\documentclass[a4paper,11pt]{article}
\usepackage{jinstpub} 
\usepackage{lineno}
\usepackage{subcaption}
\usepackage{siunitx}

\title{\boldmath Performance and Design Validation of CMS Phase-2 Pixel Modules}

\author{Giorgia Bonomelli on behalf of the CMS Tracker Group}
\affiliation{Institute for Particle and Astroparticle Physics \\
ETH Zürich,\\
Otto-Stern-Weg 5, 8093 Zürich, \\
Switzerland \\
https://orcid.org/0009-0003-0647-5103}

\emailAdd{gbonomelli@phys.ethz.ch}

\abstract{In view of the High Luminosity LHC, the current CMS tracking detector will have to be replaced during Long Shutdown 3 to cope with the higher radiation environment and to withstand an increased data rate. To prepare for the so-called CMS Phase-2 upgrade, multiple studies were carried out to characterize the pixel module design and its performance with a particular focus on the Quality Control (QC) and Assurance. For this purpose, different aspects were put together to establish a module full-performance test procedure, and novel techniques became part of the module design validation process for the full-size readout chip (CROCv1). Based on the results collected on CROCv1 prototype modules and according to the module selection criteria the community agreed on, some changes were introduced in the module design to improve the performance. This resulted in multiple prototype versions, including the production of the definitive chip (CROCv2).
\\
This study presents the quality control test flow performed, both for the dual and quad-chip module designs, on a big sample of CROCv1 prototypes and on several Kick-off and CROCv2 pre-production modules. In particular, the validation process includes measurements of the readout chip powering, sensor IV bias and open bump bonds identification. Thermal stress tests in extended temperature ranges were performed only on a subset of pixel modules to ensure the integrity of the sensor and to provide quick feedback on the quality of the bump bond connectivity after harsh temperature cycles. 
}

\begin{document}
\maketitle
\flushbottom

\section{Introduction}
\label{sec:intro}

\begin{figure}[htbp]
\centering
\includegraphics[width=.7\textwidth]{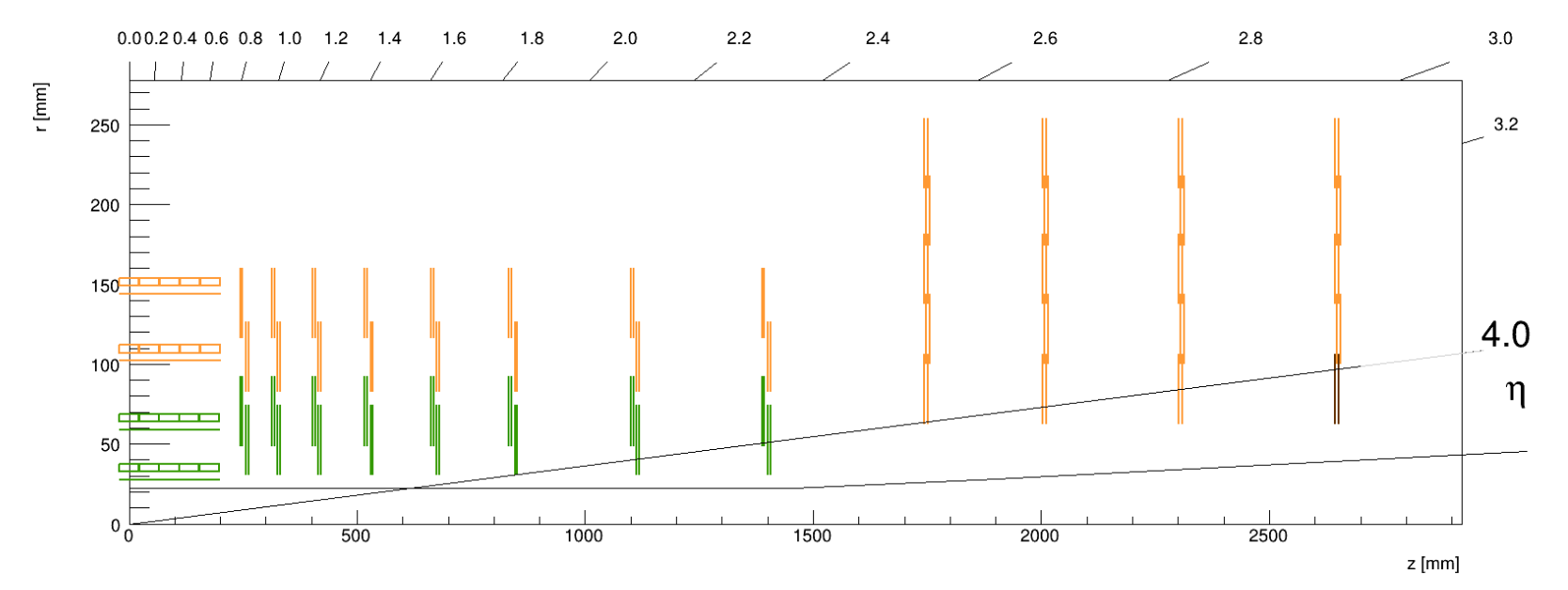}
\caption{One quarter of the CMS Inner Tracker layout in the $r-z$ plane. Green lines correspond to pixel modules with two readout chips and orange lines represent modules with four chips. \cite{Orfanelli:2780125} \label{fig:IT}}
\end{figure}

The Large Hadron Collider (LHC) at CERN is the world’s largest and highest energy particle accelerator, playing a crucial role in exploring the frontiers of high-energy physics. 
\\
To further expand the research boundaries and to enable the observation of rare processes that occur below the current sensitivity level, an upgraded machine, known as High Luminosity LHC (HL-LHC) is planned to start operation in 2030 \cite{Aberle:2749422}.
\\
During this phase the radiation levels will exceed the nominal values for which the CMS detector was designed \cite{CMS:2008xjf}. According to the CERN schedule, a rise is expected in the instantaneous luminosity with a peak of \SI{7.5e34}{\centi\meter^{-2}\second^{-1}}, and in the integrated luminosity with \SI{3000}{\femto\barn^{-1}} after ten years of operation. This will result in a significant increase in the number of interactions per bunch crossing (pileup), reaching values up to 200, approximately five times higher than the present ones of $\sim$40. In such environment, the current Pixel detector (Phase-1) \cite{CERN-LHCC-2000-016}, which of all parts of CMS is the closest to the interaction point, would be unable to operate. An entirely new and improved Inner Tracker (IT) will be installed for the Phase-2 upgrade \cite{ph2IT}, to operate at least at the same performance levels as the current detector.
\\
The CMS Inner Tracker layout for Phase-2, shown in Figure \ref{fig:IT}, consists of three different regions, with an acceptance up to $|\eta| \approx 4$. The barrel part, called TBPX, is composed of four cylindrical layers arranged in ladders. Eight small and four large double-discs per side form respectively the forward (TFPX) and the end-cap (TEPX) regions of the IT.

\section{Phase-2 Pixel Modules}
\begin{figure}[htbp]
\centering
\includegraphics[width=.45\textwidth]{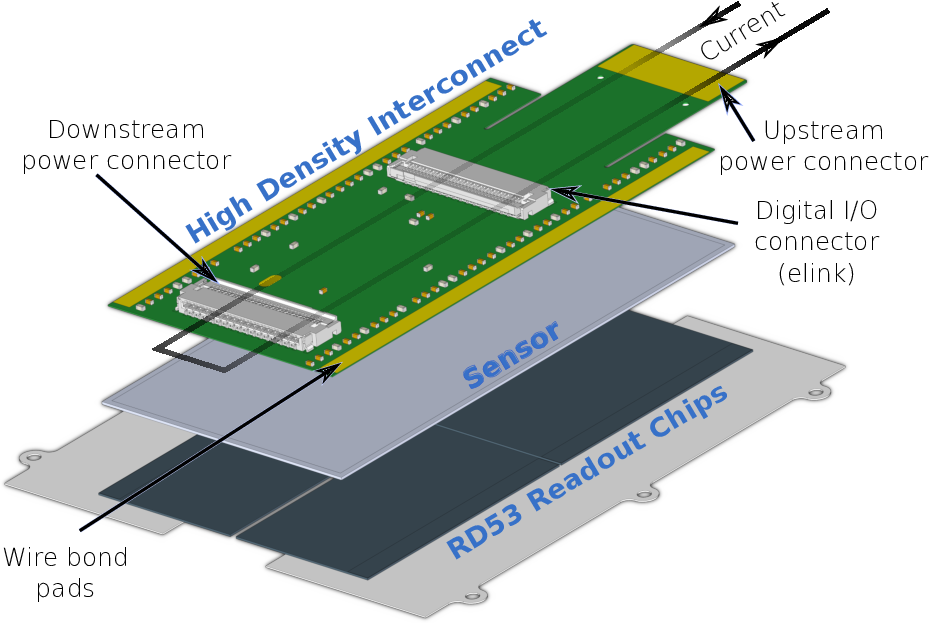}
\caption{Inner Tracker Phase-2 pixel module with four readout chips. \cite{Adam:2765909} \label{fig:module}}
\end{figure}

An IT pixel module consists of a single silicon sensor tile connected via bump bonds to one, two or four readout chips (ROCs). The sensor, which can either be planar or 3D is also glued to a High-Density Interconnect (HDI), a flexible PCB used to distribute data and power from and to the ROC (Figure \ref{fig:module}). 

To cope with the higher radiation environment and to withstand an increased particle interaction rate during the HL-LHC, a higher granularity is needed in the IT. The surface of a pixel is reduced by a factor of 6 compared to Phase-1, with the use of a silicon sensor featuring a pixel size of $25\times 100 \mu m$.
At the same time to achieve the needed logic density to handle the increased trigger rate a 65 nm CMOS technology was chosen for the readout chip. The required power consumption can only be handled using a serial powering concept \cite{Koukola:2712278}, which is a novelty for large detector systems. This scheme is designed to provide the input current to a chain of devices, each of them using a Shunt Low Drop Out (Shunt-LDO) regulator \cite{Ristic:2720556}. The voltage regulator generates the necessary voltage for the chip to operate, while the shunt circuit dissipates the current not used by the LDO into heat.

\section{Module Quality Control}
The Quality Control (QC) is a crucial step in assessing the performance of the pixel prototype modules and, based on the results, establishing the upgrade requirements for the production phase. As an iterative process, improvements and modifications can be made in the module design, resulting in multiple prototype versions (Table \ref{tab:module_versions}) until each module component is finalized. 

\begin{table}[htbp]
\centering
\caption{Different pre-production module versions.\label{tab:module_versions}}
\smallskip
\begin{tabular}{c|c}
\hline
Prototype name & Properties \\
\hline
CROCv1 & First full-size prototype chip v1, preliminary bump bonding technique \\
Kick-off & Prototype HDI design, final bump bonds technique, chip v1\\
CROCv2 & Prototype HDI design, final bump bonds technique, chip v2\\
\hline
\end{tabular}
\end{table}

\begin{figure}[htbp]
\centering
\begin{subfigure}{0.47\textwidth}
        \centering
        \raisebox{0.2cm}{\includegraphics[width=.95\textwidth]{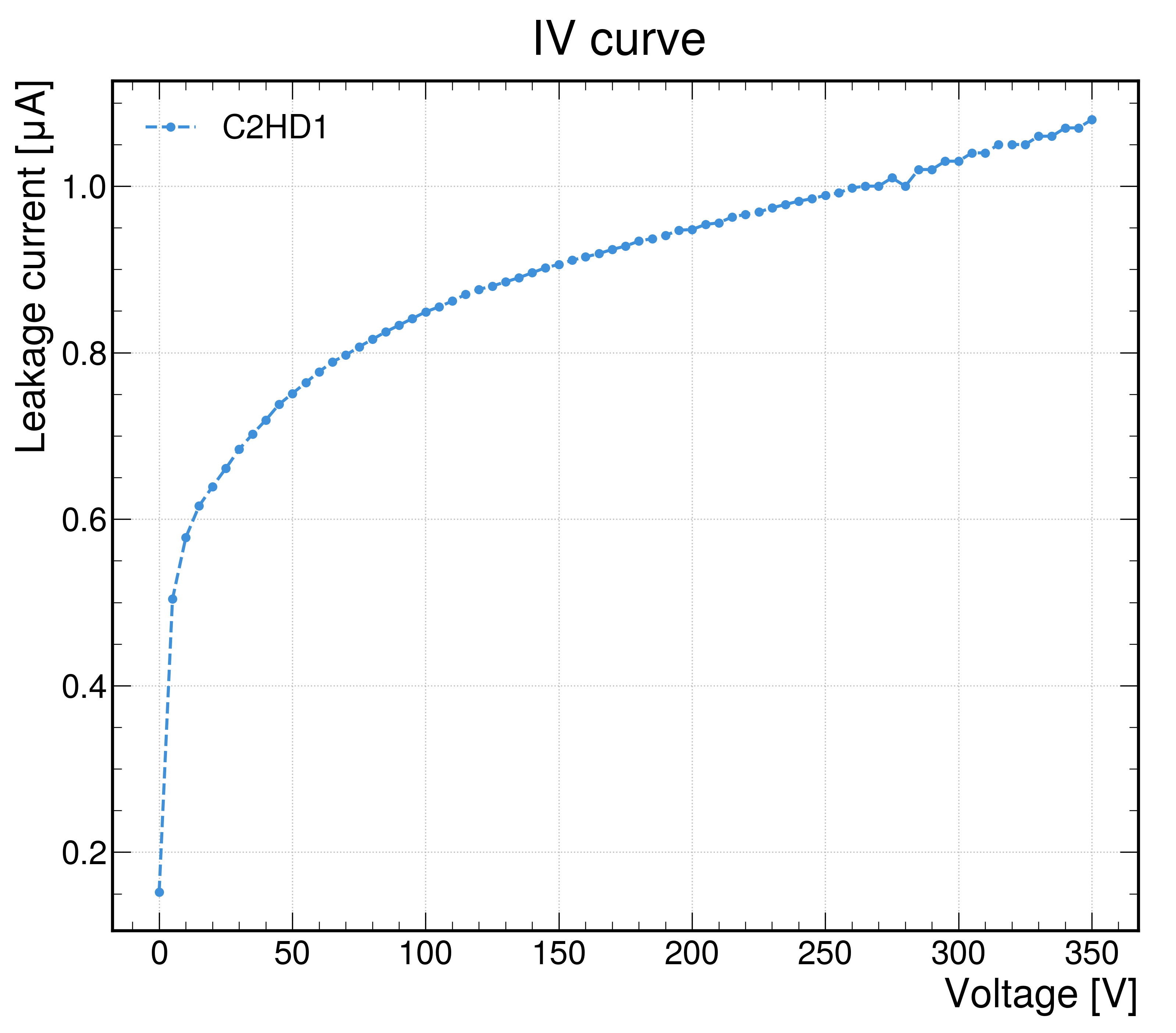}}
        \caption{Sensor IV curve of a planar module, with axes in absolute values. No breakdown is observed.}
        \label{fig:iv}
    \end{subfigure}
    \hspace{0.5cm}
    \begin{subfigure}{0.47\textwidth}
        \centering
        \includegraphics[width=1.1\textwidth]{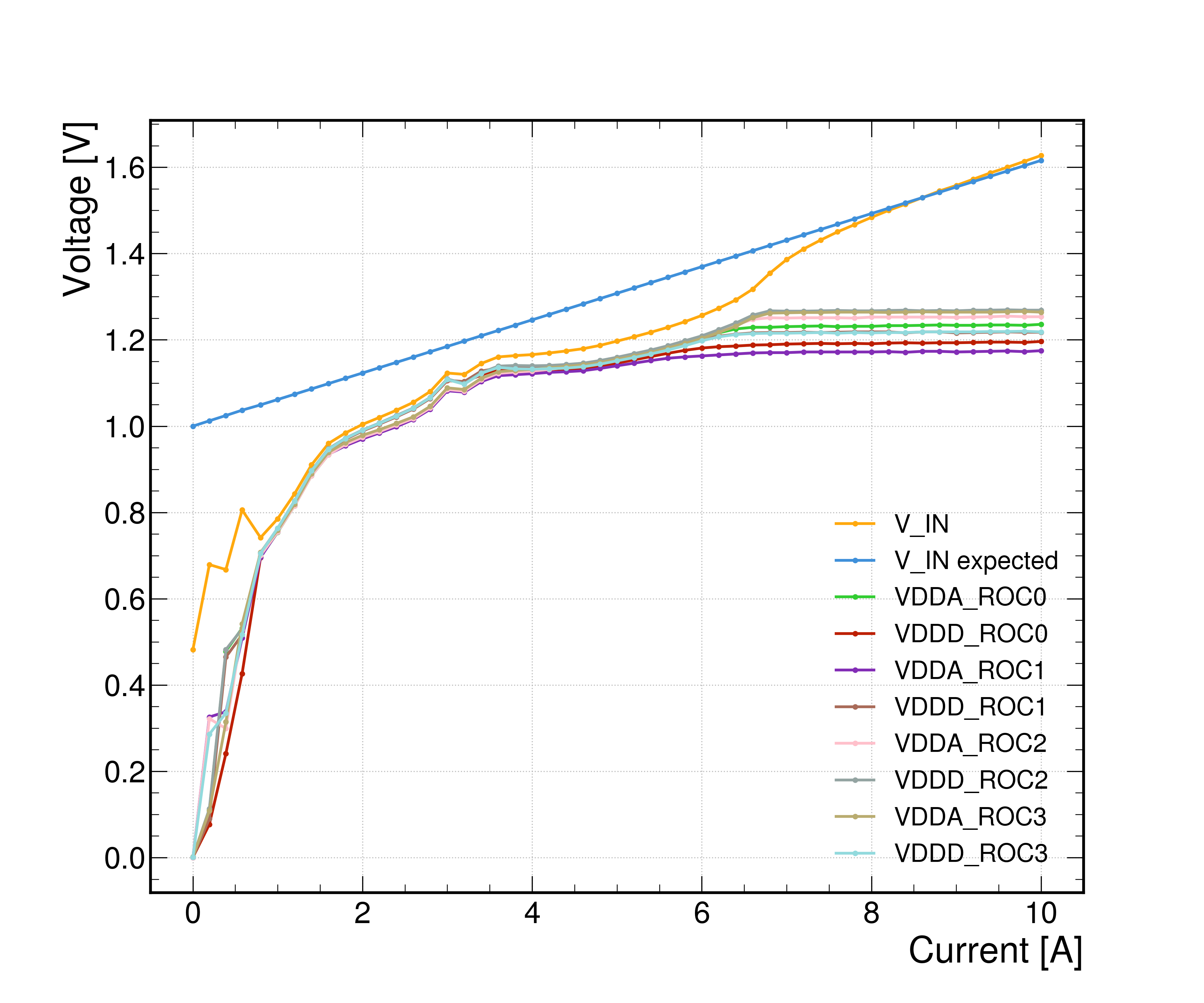}
        \caption{Shunt-LDO powering curve of a quad-chip module.}
        \label{fig:sldo}
    \end{subfigure}
    \caption{Quality Control tests on the individual components of the pixel modules, namely the sensor and the ROC.}
    \label{fig:qc}
\end{figure}

At this stage the production pixel modules are commissioned and then assembled in different test centers, where they will undergo the QC procedure to ensure they meet the requirements for installation in the future CMS IT detector.
\\ 
The Quality Control test flow has been established, optimized and automated throughout the past two years. The procedure starts with a series of tests on the individual components of a pixel module, including the ROC, the sensor and the HDI. 
\\
Among the tests that are typically performed there is the sensor IV curve. In this step the current flowing into the sensor is studied as a function of the applied bias voltage, which is increased in small steps, to identify the breakdown voltage. The IV trend is expected to be linear above full depletion and to rise exponentially as it approaches the breakdown voltage. The high voltage limits and the QC requirements are different according to the sensor type. For planar modules, tests are performed up to $-350$ V and up to $-35$ V for 3D modules. If the breakdown is not observed in these tested bias ranges, the sensor is considered good. An example is shown in Figure \ref{fig:iv}.

Next follows the powering test of the Shunt-LDO circuit of the ROC \cite{sldo}. The input, digital and analog output voltage are measured for each readout chip for different supply currents. The purpose is to verify that the operational point at which the Shunt-LDO starts providing a constant output voltage to the digital and analog domain is around 6 A for quad-chip modules (2x2) and 3 A for dual-chip modules (1x2). An example is shown in Figure \ref{fig:sldo}.

\begin{figure}[htbp]
\centering
\includegraphics[width=.85\textwidth]{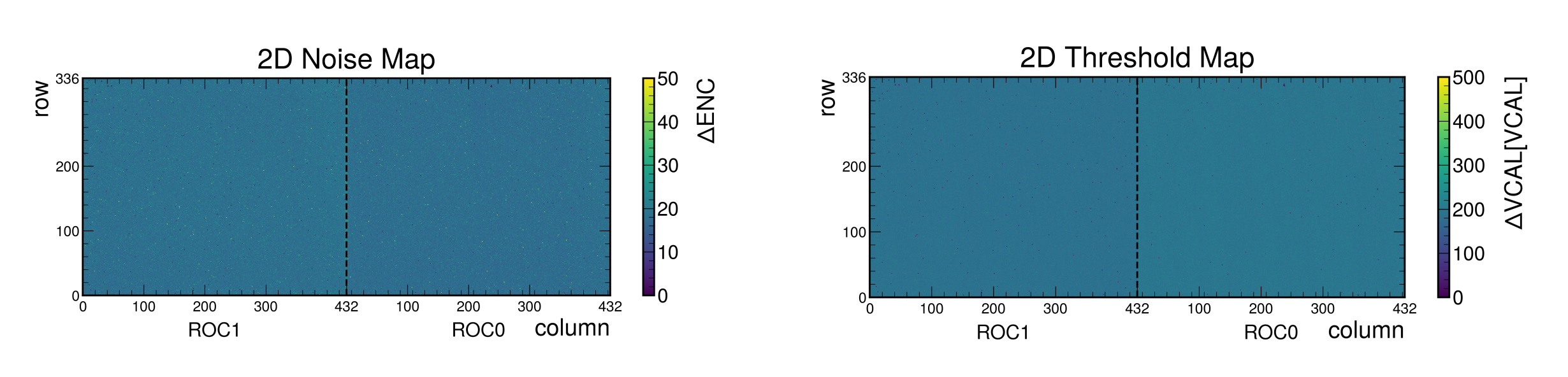}
\caption{2D noise (left) and threshold (right) map of a dual-chip module after the tuning process at low threshold. The noise distribution is in units of equivalent noise charge (ENC), while the threshold distribution in an arbitrary unit: VCAL = 5$e^{-}$. \cite{Christiansen:2898416} \label{fig:tuning}}
\end{figure}

After assembly, the modules are tested to ensure the sensor and the ROC integrity and to verify their correct functionality. This process, known as "tuning" or "calibration", begins with a quick communication test to verify the data link \cite{aurora}\cite{Orfanelli:2780125} with the readout chip. Next, an analog pulse is sent into each pixel to identify those that do not respond, which are then masked for subsequent tests. The overall threshold at which each pixel responds when it receives a hit has to be tuned to a fairly low level, usually around 1000$e^{-}$, and finely adjusted around that value. At the same time, noisy pixels that record signals without analog input are also masked. The ultimate goal of this process is to exclude non-responsive or problematic pixels and ensure a uniform response from the remaining ones when triggered. To confirm the success of the tuning procedure, the visualization of the pixel by pixel noise levels and thresholds should resemble uniform maps, as shown in in Figure \ref{fig:tuning}.
\\

After the comparison between the outcome of the QC and the selection criteria the community agreed on, all modules are graded and the test results are uploaded to a central data base.

\section{Quality Assurance}

\begin{figure}[htbp]
\centering
\begin{subfigure}{0.47\textwidth}
        \centering
        \includegraphics[width=1.12\textwidth]{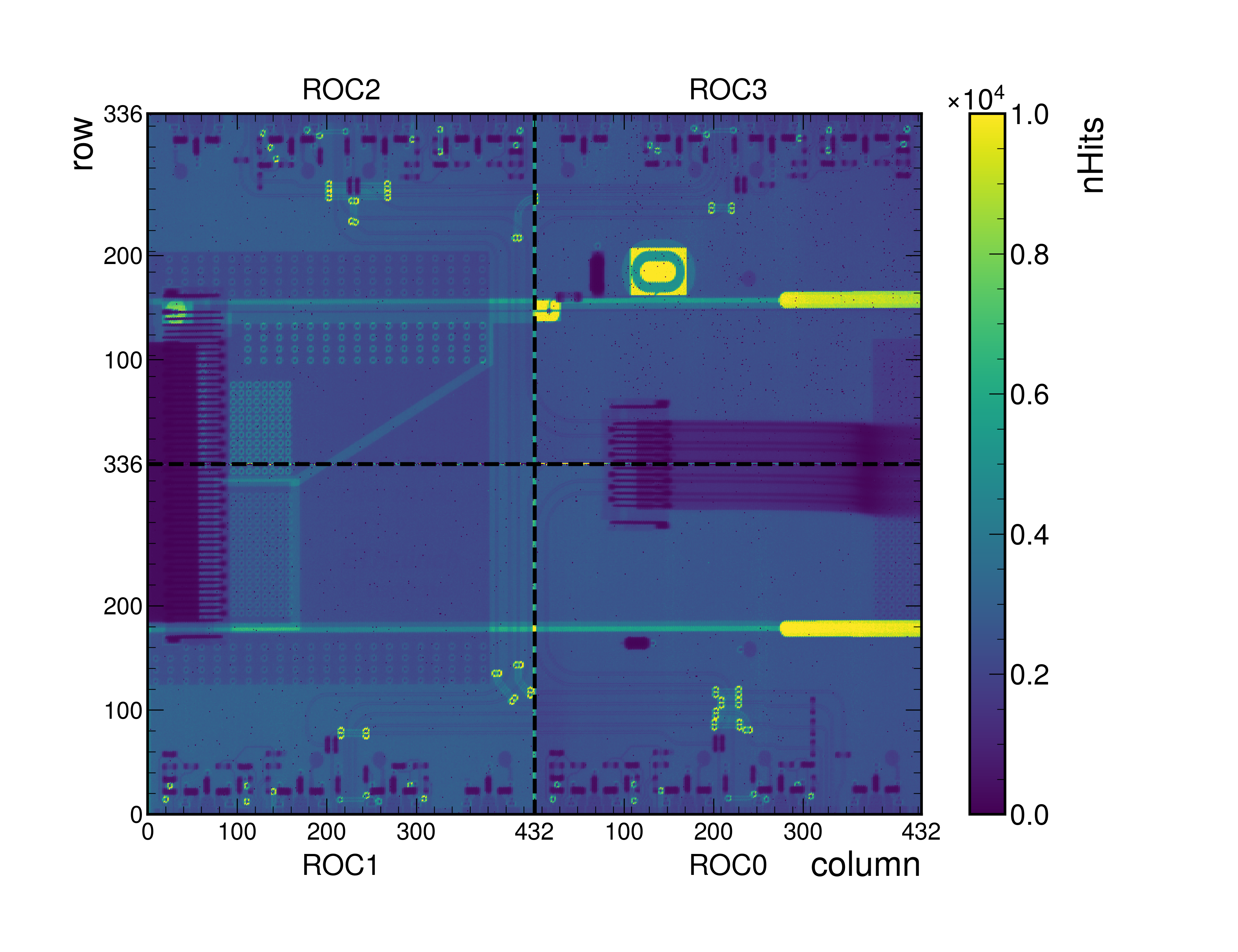}
        \caption{Occupancy map of a quad-chip module with low number of open bumps detected.}
        \label{fig:xray_map}
    \end{subfigure}
    \hspace{0.5cm}
    \begin{subfigure}{0.47\textwidth}
        \centering
        \raisebox{0.2cm}{\includegraphics[width=0.98\textwidth]{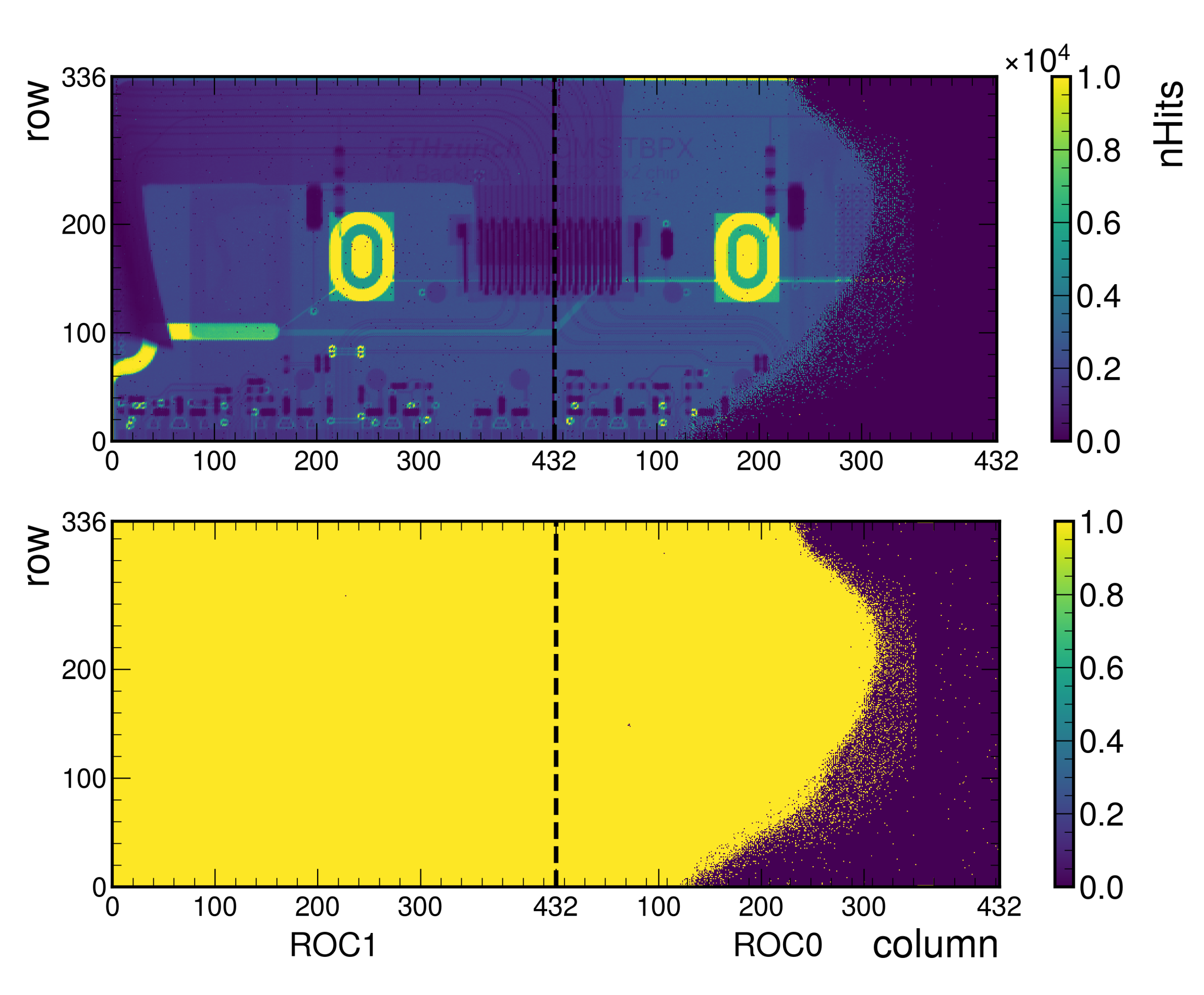}}
        \caption{Occupancy and open bumps map of a dual-chip module with high number of open bumps.}
        \label{fig:dual_xray}
    \end{subfigure}
    \caption{Occupancy maps generated with an X-ray scan for a quad and a dual-chip modules.}
    \label{fig:xray}

\end{figure}

After testing the individual module components and tuning the pixels response to the same threshold, faulty bump bond connections must be identified. Various methods can be employed for this task, but the most reliable is the radiative technique, which uses an X-ray source to quickly assess the bump quality. 
When the X-ray beam illuminates the pixel module, each pixel should register numerous hits which are recorded in an occupancy map, that also shows the HDI structures (Figure \ref{fig:xray_map}). 

\begin{figure}[htbp]
\centering
\includegraphics[width=\textwidth]{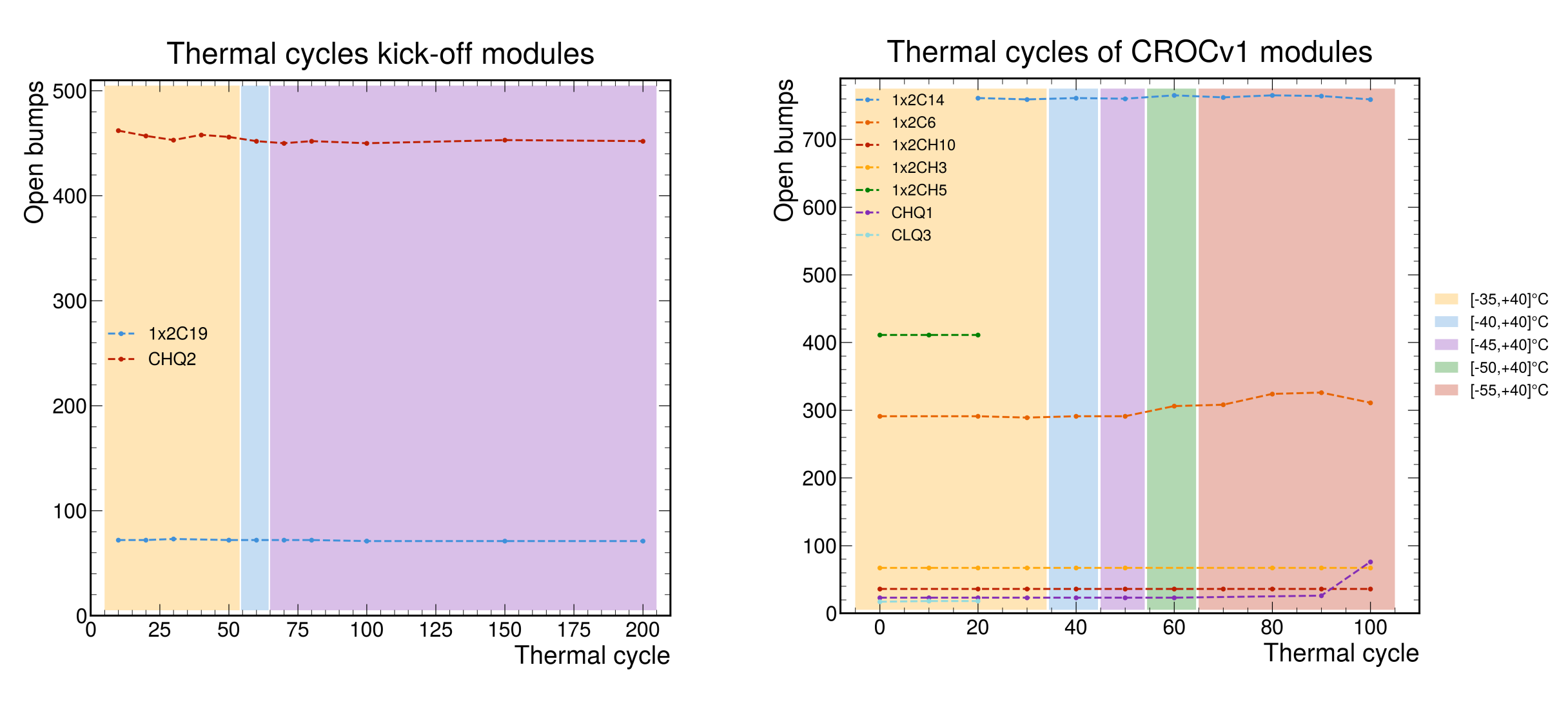}
\caption{Open bump analysis after thermal stress tests for Kick-off (left) and CROCv1 (right) prototype modules, including both 1x2 and 2x2 designs. The number of bad bump connections is measured and monitored as a function of the thermal cycles, which have been performed in various extended temperature ranges. \label{fig:thermal_cycles}}
\end{figure}

Open bumps correspond to pixels that do not produce any signal when exposed to X-rays, as the open connection either prevents signal transmission from the sensor to the readout chip or significantly reduces its efficiency.
After an accurate analysis of the occupancy map produced with this method, it is possible to identify exactly the location of the problematic or open bumps (Figure \ref{fig:dual_xray}). The total number per-chip must not exceed 600 to stay within the upgrade requirements, which corresponds to $\sim 0.4 \%$ of the pixel matrix. 
\\
If a pixel module passed the selection criteria based on the results of the QC procedure and the X-ray tests, it will be sent for coating with a spark protection layer. Once back, it will undergo the full performance test once more and, in addition, thermal cycles. These thermal stress tests are applied for the first time to CMS pixel modules for Phase-2 in this context and they are meant to test the resistance of the bump bonds under large temperature gradients. In addition, only a subset of modules will be exposed to extended temperature ranges, in order to test the design limits. To monitor the bump bond status and the formation of new clusters of open connections throughout the thermal stress tests, the X-ray method is used after each subset of ten cycles. 
As depicted in Figure \ref{fig:thermal_cycles}, most modules subjected thermal stress test did not show any significant increase in the number of open bonds, even when the temperature range was significantly enlarged. These results were crucial in validating the design and the bump bonding technique for the production modules.

\section{Results}
The full-performance test strategy was newly developed and improved on the first CROCv1 prototype modules, and used to define the requirements for the Quality Control results for the Phase-2 upgrade.
Through the Quality Control procedure, various design changes were implemented to optimize the pixel modules performance, starting with the first full-size prototype chip CROCv1, progressing through the Kick-off modules, and leading to the final prototypes before production CROCv2 (Table \ref{tab:module_versions}).
A good modules yield was recorded for each pre-production design version, not only in terms of preliminary tests at the sensor, readout chip and full module level, but also concerning the stringent open bumps requirement. On a total of 98 tested chips, only 5 did not meet the stringent requirements imposed for the open bumps, bringing a yield of $\sim 94 \%$.
All the pre-series prototypes, included the closest to the final production design CROCv2, were qualified against the Quality Control requirements during the full-performance tests and withstood thermal stress tests without showing no new cluster of open bumps.

\section{Conclusions}
Several quad and dual-chip prototypes for the Phase-2 upgrade were assembled and fully tested in laboratory to verify their correct operation and to evaluate their performance.
All pixel modules behave as expected in terms of Shunt-LDO powering of the readout chip, sensor IV curve, and threshold and noise tuning. These measurements, together with the extensive thermal cycles campaign, were crucial in validating the individual components and the full pixel module design. The successful results achieved with the pre-production qualification process confirm that, even after some minor design changes, the latest version of the readout chip, CROCv2, and the bump bonding technique still fulfill the requirements for Phase-2. With these newly revised module components, the commissioning and production phase for the final CMS Pixel module layout has now started. Once the  sensors and readout chips become available, hundreds of modules will be assembled at different production centers and undergo the full-performance testing before installation in the CMS detector. 


\bibliographystyle{unsrt} 
\bibliography{biblio}     

\end{document}